\documentclass[aps,prb,citeautoscript,twocolumn,superscriptaddress,footinbib,10pt]{revtex4-2}
\usepackage[utf8]{inputenc}
\usepackage{amsmath}
\usepackage{amssymb}
\usepackage{mathrsfs}
\usepackage{color}
\usepackage{graphicx,color,colortbl}
\usepackage{rotating,array,tabularx,booktabs}
\usepackage{varwidth,xcolor}
\usepackage{placeins}
\usepackage{siunitx}
\usepackage[capitalize]{cleveref}
\usepackage{soul}  
\usepackage{dsfont}

\DeclareGraphicsExtensions{.pdf,.PDF,.png,.PNG}

\newcommand{\dif}{\mathrm{d}}%
\newcommand{\Eins}{\underline{1}}%

\newcommand{\ii}{\mathrm{i}}%
\newcommand{\Nabla}{\boldsymbol{\nabla}}%
\newcommand{\norm}[1]{\lVert#1\rVert}%
\newcommand{\ZT}[1]{\textquotedblleft#1\textquotedblright}%

\newcommand{\Pecn}{P\'{e}clet number }
\newcommand{\Pecns}{P\'{e}clet numbers }

\newcommand{\Pecnc}{P\'{e}clet number, }
\newcommand{\Pecnsc}{P\'{e}clet numbers, }
\newcommand{\Pecnsp}{P\'{e}clet numbers. }

\newcommand{\tauLJ}{\tau_\mathrm{LJ}}
\newcommand{\urep}{q}

\newcolumntype{Y}{>{\centering\arraybackslash}X}%
\begin{document}
\title{Pair-distribution function of active Brownian spheres in three spatial dimensions: simulation results and analytical representation}

\author{Stephan Br\"oker}
\affiliation{Institute of Theoretical Physics, Center for Soft Nanoscience, University of M\"unster, 48149 M\"unster, Germany}

\author{Michael te Vrugt}
\affiliation{Institute of Theoretical Physics, Center for Soft Nanoscience, University of M\"unster, 48149 M\"unster, Germany}

\author{Julian Jeggle}
\affiliation{Institute of Theoretical Physics, Center for Soft Nanoscience, University of M\"unster, 48149 M\"unster, Germany}

\author{Joakim Stenhammar}
\affiliation{Division of Physical Chemistry, Lund University, 221 00 Lund, Sweden}

\author{Raphael Wittkowski}
\email[Corresponding author: ]{raphael.wittkowski@uni-muenster.de}
\affiliation{Institute of Theoretical Physics, Center for Soft Nanoscience, University of M\"unster, 48149 M\"unster, Germany}

\begin{abstract}
The pair-distribution function, which provides information about correlations in a system of interacting particles, is one of the key objects of theoretical soft matter physics. In particular, it allows for microscopic insights into the phase behavior of active particles. While this function is by now well studied for two-dimensional active matter systems, the more complex and more realistic case of three-dimensional systems is not well understood by now. In this work, we analyze the full pair-distribution function of spherical active Brownian particles interacting via a Weeks-Chandler-Andersen potential in three spatial dimensions using Brownian dynamics simulations. Besides extracting the structure of the pair-distribution function from the simulations, we obtain an analytical representation for this function, parametrized by activity and concentration, which takes into account the symmetries of a homogeneous stationary state. Our results are useful as input to quantitative models of active Brownian particles and advance our understanding of the microstructure in dense active fluids. 
\end{abstract}
\maketitle
\section{Introduction}
The study of active soft matter \cite{BechingerdLLRVV2016,MarchettiJRLPRS2013} has been one of the most rapidly growing fields of physics in the past decade. Active particles convert energy into directed motion, as a consequence of which active matter systems are permanently driven out of equilibrium. This gives rise to a broad range of phenomena that are not possible in equilibrium (passive) systems. Examples of active matter include biological organisms like bacteria \cite{DrescherKD2011,berg2008,PetroffAXLA2015}, birds \cite{BialekCG2012}, or fish \cite{BalleriniCC2008}, but also synthetic objects like self-propelled catalytic Janus particles \cite{WaltherM2008}. 

Rather than using a detailed description of living organisms, theoretical studies of the collective dynamics of active matter typically start from simple models. Among the most important one of these is the \textit{active Brownian particle} (ABP), which exhibits a self-propulsion force with constant magnitude in a direction $\boldsymbol{\widehat{u}}$ that changes via rotational diffusion. ABPs exhibit interesting collective dynamics, with one of the most notable and widely studied phenomena being motility-induced phase separation (MIPS) \cite{TailleurC2008}. Here, a system consisting of particles with purely repulsive interactions separates into a  dilute gas phase and a dense cluster phase. Most studies of MIPS and the collective dynamics of ABPs focus on the two-dimensional case \cite{CatesT2015, BialkeLS2013, StenhammarWMC2015, CatesT2013, ButtinoniBKLBS2013, FilyM2012, FilyHM2014, RednerHB2013, WittkowskiTSAMC2014, BialkeSLS2015, BlaschkeMMZ2016,DigregorioLSCGP2018,TheersWQWG2018,FischerCS2019, KetaR2019,NavarroF2015,JeggleSW2020}. 
The three-dimensional case has also attracted some attention \cite{StenhammarMAC2014,WysockiWG2014,SiebertLSV2017,DasGW2018,AlarconP2013,NieCPDNC2020a,PrymidisPDF2016,PrymidisSF2015,FarageKB2015,ReinS2016}, but is, due do its increased computational complexity, less well understood. 
\\

A central quantity of classical many-body physics is the pair-distribution function $g$, which determines the probability of finding two particles in a particular configuration. The importance of this function results, among other things, from the fact that it is used in microscopic derivations of field-theoretical models \cite{WittkowskiSC2017,BickmannW2020a, BickmannW2020b,teVrugtBW2022,BroekerBtVCW2022}. For equilibrium systems, the pair-distribution function is well understood. It can be approximated analytically using liquid integral theory \cite{GrayG1984,MoritaH1960, Percus1964, Stell1964, HansenMD2009} and has been thoroughly studied both in experiments \cite{vanBlaaderenW1995,CarbajalTinocoCRAL1996, Hughes2010,IacovellaRGS2010,ThorneyworkRAD2014} and computer simulations \cite{IacovellaRGS2010, AllenT2017}. 
These results, however, do not generally carry over to the active case: Since ABPs are far from thermodynamic equilibrium, their properties cannot be related to their bare interaction potentials alone. Instead, one needs to also take into account the effect of self-propulsion on the local properties. At the same time, the pair-distribution function has been found to be highly useful in understanding the microscopic origin of MIPS \cite{BialkeLS2013}. 

The pair-distribution function of active particles has therefore attracted an increasing amount of interest in recent years \cite{WittkowskiSC2017,BialkeLS2013,SchwarzendahlM2018,HaertelRS2018,PessotLM2018, JeggleSW2020}. In particular, it has been investigated how this function depends on the activity and packing density of the particles, both in single-component systems \cite{BialkeLS2013} and in mixtures of active and passive particles \cite{WittkowskiSC2017}. Moreover, \citet{HaertelRS2018} have analyzed the three-body distribution and the full pair-distribution function of ABPs in two spatial dimensions. They also obtained an analytical expression for a reduced form of the pair-distribution function. \citet{SchwarzendahlM2018} have studied the influence of hydrodynamic interactions on the pair-distribution function, and also investigated the (not fully orientation resolved) pair-distribution function of ABPs in three spatial dimensions. The orientational ordering and collective behaviour of pushers and pullers was investigated in Ref.\ \cite{PessotLM2018}. Additionally, an approximate expression valid in the slow- and fast-swimming limits for the pair-distribution function of ABPs was derived by \citet{DhontPB2021motility}. The validity of this expression, however, is limited to packing densities below $0.1$ \cite{DhontPB2021motility}.
Finally, a fully orientation-resolved pair-distribution function for a wide range of activities and packing densities in two spatial dimensions has been published recently by us \cite{JeggleSW2020}, along with a software package that allows to reproduce this function \cite{JeggleSW2019b}. By comparing with the results from Refs.\ \cite{WittkowskiSC2017,BickmannW2020a} we furthermore observe that the availability of such a fully orientation-resolved distribution significantly improves theoretical predictions for the spinodal of MIPS.

However, until now there is no systematic analysis of the full orientation-resolved pair-distribution function of spherical ABPs for a wide range of activities and
packing densities in three spatial dimensions. To close this gap, we study in this work the fully orientation-resolved pair-distribution function using Brownian dynamics simulations and examine its dependence on all relevant parameters for homogeneous stationary states. We also provide an analytical expression for the product of the interaction force and the pair-distribution function.
This analytical expression has already been used in Ref. \cite{BickmannW2020b} to find the spinodal and critical point of a system of ABPs. 

This article is structured as follows. We explain our methodology and give an overview of the simulation details in Sec.\ \ref{sec:methods}.
In Sec.\ \ref{sec:results}, we present a high-resolution state diagram, the fully orientation-resolved pair-distribution function, and an analytical expression for the product of this function and the interaction force. We conclude in Sec.\ \ref{sec:conclusions}.
\section{\label{sec:methods}Methods}
To analyze the pair-distribution function of spherical ABPs, we carried out Brownian dynamics simulations using a modified version of the software package LAMMPS \cite{Thompson2022lammps}.
\subsection{Model and simulation details\label{subsec:Modandsim}}
We study the dynamics of $N$ spherical ABPs, described by overdamped Langevin equations \cite{BialkeLS2013,RednerHB2013,StenhammarMAC2014,JeggleSW2020}. For the translational motion, they are given by
\begin{align}
\dot{\boldsymbol{r}}_i &= - \frac{D_\mathrm{t}}{k_\mathrm{B} T} \sum^{N}_{\begin{subarray}{c}j=1\\j \neq i\end{subarray}} \nabla_{\boldsymbol{r}_i} U(\norm{\boldsymbol{r}_i- \boldsymbol{r}_j  } )  +
v_0
\, \widehat{\boldsymbol{u}}_i + \boldsymbol{f}_{\mathrm{t},i}
\end{align}
with the position $\boldsymbol{r}_i$ of the $i$-th particle, the time $t$, the translational diffusion coefficient of passive spherical particles $D_{\mathrm{t}} = k_\mathrm{B} T/(3 \pi \sigma \eta)$ resulting from the Stokes-Einstein relation, the Boltzmann constant $k_\mathrm{B}$, the absolute temperature $T$, the particle diameter $\sigma$, the dynamical viscosity $\eta$, the interaction potential $U(r)$, the self-propulsion speed $v_0$,
the orientation of the $i$-th particle $\boldsymbol{\widehat{u}}_i$, and the noise term $\boldsymbol{f}_{\mathrm{t},i}(t)$.
For $U$, we use the Weeks-Chandler-Andersen potential \cite{WeeksCA1971}, i.e., a truncated and shifted Lennard-Jones potential. This purely repulsive potential reads
\begin{equation}
U(r)=
\begin{cases}
4\varepsilon \Big( \big( \frac{\sigma}{r} \big)^{12} - \left( \frac{\sigma}{r} \right)^{6} \Big) + \varepsilon & \mbox{if } r \leq 2^{1/6} \sigma, \\
0 & \mbox{else}
\end{cases}
\label{eq:wca}%
\end{equation}
with the interaction strength $\varepsilon$.
The noise term $\boldsymbol{f}_{\mathrm{t},i}(t)$ is modeled using Gaussian white noise with the properties $\langle \boldsymbol{f}_{\mathrm{t},i}(t) \otimes \boldsymbol{f}_{\mathrm{t},j}(t^{\prime}) \rangle = 2 D_{\mathrm{t}} \,  \delta_{i,j} \, \delta(t-t^{\prime}) \Eins$, where $\otimes$ is the dyadic product and $\Eins$ the identity matrix.
To quantify the ratio between active and thermal forces, we use (as is common) the \Pecn $\mathrm{Pe} = v_0 \sigma / D_\mathrm{t}$. \\
The rotational motion of the particle is given by
\begin{align}
\dot{\widehat{\boldsymbol{u}}}_i &=\widehat{\boldsymbol{u}}_i \times \boldsymbol{f}_{\mathrm{r},i} \,
\end{align}
with $\boldsymbol{f}_{\mathrm{r},i}(t)$ being a noise term that is characterized as Gaussian white noise with the properties $\langle \boldsymbol{f}_{\mathrm{r},i}(t) \otimes \boldsymbol{f}_{\mathrm{r},j}(t^{\prime})\rangle = 2 D_{\mathrm{r}}  \delta_{i,j}  \delta (t-t^{\prime}) \Eins$. 
Here, $D_\mathrm{r}$ is the rotational diffusion coefficient that for spherical particles in a viscous fluid is given by the Stokes-Einstein-Debye relation $ D_{\mathrm{r}} = k_\mathrm{B} T / (\pi \sigma^3 \eta) = 3 D_\mathrm{t}/\sigma^2$.
The swimming speed of a particle is $v_0 = 24 \sigma/ \tauLJ =  24 D_\mathrm{t} \varepsilon/(\sigma k_\mathrm{B} T )$, such that the repulsive force of the potential at $r=\sigma$ is equal to the force corresponding to self-propulsion. We vary the \Pecn by changing the temperature \cite{StenhammarMAC2014, JeggleSW2020}: as the temperature diverges for small \Pecnsc we vary $\mathrm{Pe}$ between $50$ and $500$. We use the Lennard-Jones units $\tauLJ= \sigma^2/(\varepsilon \beta D_\mathrm{t})$, $\sigma$, and $\varepsilon$ as units of time, length, and energy, respectively.
\\
Besides the \Pecnc we vary the average packing density $\Phi_0 = \pi \sigma^3 N/(6V)$ (where $V$ is the volume of the domain) by changing the particle number $N$. The domain is a cubic box with periodic boundary conditions and side length $30\sigma$. Varying $\Phi_0$ from $0.05$ to $0.7$ corresponds to $N \in [2578,36096]$.
The particles were initially placed on a hexagonal grid to prevent strong overlapping. After an initial simulation time of $150 \tauLJ$ at a low \Pecn ($\mathrm{Pe} = 50$), the positions of the particles are relaxed to the steady state distribution. Then, the pair-distribution function was analyzed by sampling the system configuration at regular time intervals and binning the relative configuration for all particle pairs with a distance less than $10\sigma$.
Aiming for an average of $500$ entries in the bins for the distance $r=\sigma$, we chose the simulation time according to the number of particles in the simulation. Thus, the simulation time scaled quadratically with the inverse of the number of particles and varied between $2.4\cdot 10^4 \, \tauLJ$ and $4.7 \cdot 10^6 \,\tauLJ$ depending on $\Phi_0$.
The time step size was $5 \cdot 10^{-5} \tauLJ$ and the time between samples was chosen such that a noninteracting self-propelled particle would be displaced by twice its diameter.
\begin{figure}[htb]
\centering
\includegraphics[]{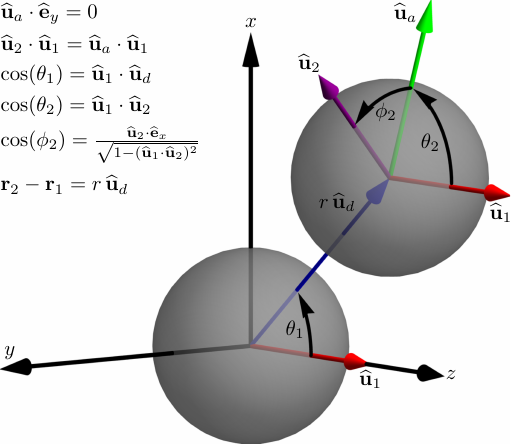}%
\caption{\label{fig1}
Parametrization of the fully orientation-resolved pair-distribution function. The center of the coordinate system coincides with the center of the first particle and the $z$-axis is parallel to the orientation $\widehat{\boldsymbol{u}}_1$ of the first particle. The angle $\theta_1$ is the angle between the orientation of the first particle and the vector $\boldsymbol{u}_d = r \widehat{\boldsymbol{u}}_d$ that connects the two particles, $r$ the distance between the particles, and the angles $\theta_2$ and $\phi_2$ correspond to the polar and azimuthal angle, respectively, of the second particle's orientation $\widehat{\boldsymbol{u}}_2$ relative to that of the first particle.
The vector $\widehat{\boldsymbol{u}}_a$ is a supportive vector in the $x$-$z$-plane that has the same polar angle as $\widehat{\boldsymbol{u}}_2$.}
\end{figure}

To distinguish between a system undergoing phase separation and a homogeneous system, we used the characteristic length $L_\mathrm{c}$. This length quantifies density inhomogeneities and can therefore be used as a measure for phase separation \cite{StenhammarMAC2014}. It is defined as \cite{StenhammarMAC2014}
\begin{equation}
  L_\mathrm{c}(t) = 2 \pi \frac{\int_{2 \pi /L}^{k_\mathrm{cut}} \! \dif k \, S(k,t) }{\int_{2 \pi /L}^{k_\mathrm{cut}} \! \dif k \, k \, S(k,t) } \, . \label{equ:char_length}
\end{equation}
Here, $S(k,t)$ is the structure factor \cite{BiniossekLVS2018}
\begin{equation}
    S(k,t) = \frac{1}{N} \Bigg\langle \sum_{i,j=1}^N \exp(-\ii \boldsymbol{k}\cdot(\boldsymbol{r}_i(t) -\boldsymbol{r}_j(t)  )) \Bigg\rangle,
\end{equation}
$\langle \cdot \rangle$ the average over the stationary state, $L$ the domain size, $k_\mathrm{cut}$ a cutoff wavelength, and $k=\norm{\boldsymbol{k}}$ the norm of the wave vector $\boldsymbol{k}$. We here chose $k_\mathrm{cut}=\pi$, which approximately coincides with the first minimum of $S(k,t)$. A high characteristic length corresponds to a high degree of spatial order, i.e., to phase separation, which was confirmed by visual inspection. 
The characteristic length was sampled with a time resolution of $0.42\tauLJ$ for a total simulation time between $5 \cdot 10^3\tauLJ$ and $1.5\cdot10^2\tauLJ$ scaling inversely with the packing density.
At the beginning of each simulation, an additional simulation time of $100\tauLJ$ was added to allow for relaxation.
For each parameter combination in the state diagram (see below), six simulations were performed. The characteristic length was first averaged over the last $21 \tauLJ$, i.e., $50$ samples, of each simulation, and then averaged over the different simulations.
Due to the metastability of the homogeneous state in the binodal region, it is sensible to perform multiple simulations starting from different initial conditions for each parameter combination. Typically, the rapid change in characteristic length marking the spontaneous transition between homogeneous and clustered state happens on the timescale of a few $\tauLJ$ within the first $20 \tauLJ$ of the simulation. As the characteristic length remains virtually constant after this, the comparingly short simulation times were sufficient.
Performing multiple simulations for each parameter combination, we confirmed the stability of the homogeneous distribution for parameter combinations close to the state boundary that will be used to examine the pair-distribution function.
\subsection{\label{subsec:parametrization}Parametrization of the pair-distribution function}
Let $P(\{\boldsymbol{r}_i\},\{\boldsymbol{\widehat{{u}}}_i\},t)$ be the probability that the system is, at time $t$, in the microscopic configuration specified by the coordinates $\boldsymbol{r}_i$ and $\boldsymbol{\widehat{{u}}}_i$. In this case, we can define the $n$-particle density as 
\begin{equation}
\begin{split}
&\varrho^{(n)}(\boldsymbol{r}_1,\dotsc, \boldsymbol{r}_n, \boldsymbol{\widehat{{u}}}_1,\dotsc,\boldsymbol{\widehat{{u}}}_n, t) 
\\&= \frac{N!}{(N-n)!}\Bigg( \prod\limits_{i=n+1}^N \int_{\mathds{R}^3}\!\!\!\!\mathrm{d}^3 r_i \int_{\mathbb{S}_2}^{}\!\!\!\!\mathrm{d}^2 u_i \Bigg) P(\{\boldsymbol{r}_i\},\{\boldsymbol{\widehat{{u}}}_i\},t),
\end{split}
\label{eqn:ProjectionNParticleDensity}
\end{equation}
where $\mathbb{S}_2$ is the unit sphere in three spatial dimensions. This allows to define the pair-distribution function as \cite{HansenMD2009}
\begin{equation}
g(\boldsymbol{r}_1, \boldsymbol{r}_2, \boldsymbol{\widehat{{u}}}_1, \boldsymbol{\widehat{{u}}}_2, t)=\frac{\varrho^{(2)}(\boldsymbol{r}_1, \boldsymbol{r}_2, \boldsymbol{\widehat{{u}}}_1, \boldsymbol{\widehat{{u}}}_2, t)}{\varrho(\boldsymbol{r}_1, \boldsymbol{\widehat{{u}}}_1, t)\varrho(\boldsymbol{r}_2, \boldsymbol{\widehat{{u}}}_2, t)}
\label{eq:pairdistribution}
\end{equation}
with $\varrho=\varrho^{(1)}$. The two-body density $\varrho^{(2)}(\boldsymbol{r}_1, \boldsymbol{r}_2, \boldsymbol{\widehat{{u}}}_1, \boldsymbol{\widehat{{u}}}_2, t)$ gives the probability of finding one particle with orientation $\boldsymbol{\widehat{{u}}}_1$ at position $\boldsymbol{r}_1$ and another particle with orientation $\boldsymbol{\widehat{{u}}}_2$ at position $\boldsymbol{r}_2$ at time $t$ multiplied by $N(N-1)$, and the one-body density $\varrho(\boldsymbol{r}_1, \boldsymbol{\widehat{{u}}}_1,t)$ gives the probability of finding one particle with orientation $\boldsymbol{\widehat{{u}}}_1$ at position $\boldsymbol{r}_1$ at time $t$ multiplied by $N$. If we use now the definition \cite{Myrvold2021}
\begin{equation}
P(A|B)=\frac{P(A\cap B)}{P(B)}    
\end{equation}
of the conditional probability of event $A$ given an event $B$, we can see that the product $\varrho(\boldsymbol{r}_2, \boldsymbol{\widehat{{u}}}_2,t)g(\boldsymbol{r}_1, \boldsymbol{r}_2, \boldsymbol{\widehat{{u}}}_1, \boldsymbol{\widehat{{u}}}_2, t)/(N-1)$ is simply the conditional probability of finding, at time $t$, a particle with orientation $\boldsymbol{\widehat{{u}}}_2$ at position $\boldsymbol{r}_2$ given that another particle with orientation $\boldsymbol{\widehat{{u}}}_1$ is at position $\boldsymbol{r}_1$ \cite{WeberS2012}. Since we focus on spatially homogeneous one-particle distributions in this work, we will, with a slight abuse of terminology, sometimes simply refer to $g$ as \ZT{probability}. Note, however, that strictly speaking $g$ is not a probability, but proportional to a conditional probability (this conditional probability is determined by $g$ once $\varrho$ is fixed).

The function $g$ will take different forms for different activities or packing densities. Thus, in general, the full pair-distribution function $  g(\mathrm{Pe}, \Phi_0;\boldsymbol{r}_1,\boldsymbol{r}_2, \boldsymbol{\widehat{u}}_1,\boldsymbol{\widehat{u}}_2,t)$ depends on the \Pecn Pe, the packing density $\Phi_0$, the position and orientation of both particles, and the time $t$. 
To simplify the pair-distribution function, we assume a stationary and homogeneous system. This allows us to drop the time dependence of $g$ and to reduce the dependency on the absolute positions of the two particles to one on their relative positions. In total, the pair-distribution function $g$ then depends only on the \Pecn Pe, the packing density $\Phi_0$, the particles' relative position $\boldsymbol{{r}}_2-\boldsymbol{{r}}_1=\boldsymbol{{r}} = r \boldsymbol{\widehat{u}_d}$ and the orientation $\boldsymbol{\widehat{u}}_i$ of each particle:
\begin{equation}
  g(\mathrm{Pe}, \Phi_0;r,\boldsymbol{\widehat{u}}_d, \boldsymbol{\widehat{u}}_1,\boldsymbol{\widehat{u}}_2)\, .
\end{equation}
For the sake of brevity, we omit the explicit dependency on Pe and $\Phi_0$ in our notation for the rest of this section.
We can also exploit the isotropy of the system to eliminate some orientational dependencies of $g$. For this, we first define a coordinate system with its origin at the center of the first particle and the $z$-axis aligned with its orientation. Furthermore, we fix the $x$-axis such that both particles lie in the $x$-$z$ plane with the second particle at a positive $x$ coordinate as shown in Fig.\,\ref{fig1}. 
Thus the orientation of the first particle is fixed.
The relative position vector $r \,  \boldsymbol{\widehat{u}}_d$ is now fixed to the $x$-$z$-plane and can be defined by the interparticle distance $r$ and the angle $\theta_1$ between $\boldsymbol{\widehat{u}}_1$ and $\boldsymbol{\widehat{u}}_d$.
The orientation of the second particle can be expressed via two angles.
We choose the polar angle $\theta_2$ and the azimuthal angle $\phi_2$ as shown in Fig.\ \ref{fig1}.
Thus the pair-distribution function only depends on four variables, i.e., $g=g(r,\theta_1, \theta_2,\phi_2)$.
The distance $r$ can furthermore be calculated via
\begin{equation}
  r = \norm{ \boldsymbol{r_d}}= \norm{ \boldsymbol{r}_1- \boldsymbol{r}_2}
\end{equation}
and the angles $\theta_1$ and $\theta_2$ are given by
\begin{align}
  \theta_1 &= \arccos (\boldsymbol{\widehat{u}}_1 \cdot \boldsymbol{\widehat{u}}_d), \\
  \theta_2 &= \arccos (\boldsymbol{\widehat{u}}_1 \cdot \boldsymbol{\widehat{u}}_2).
\end{align}
The unit vector in $x$-direction $\boldsymbol{\widehat{e}}_x$ can be obtained via the cross product of the unit vector in $y$-direction $\boldsymbol{\widehat{e}}_y$ and $\boldsymbol{\widehat{u}}_1$, while
$\boldsymbol{\widehat{e}}_y$ is obtained by calculating $\boldsymbol{\widehat{u}}_1 \times \boldsymbol{\widehat{u}}_d$ and normalizing the resulting vector via division by $\sin(\theta_1)$:
\begin{equation}
\boldsymbol{\widehat{e}}_x = \boldsymbol{\widehat{e}}_y \times \boldsymbol{\widehat{u}}_1 =
\frac{\boldsymbol{\widehat{u}}_1 \times \boldsymbol{\widehat{u}}_d}{\sin(\theta_1)} \times \boldsymbol{\widehat{u}}_1
=\frac{\boldsymbol{\widehat{u}}_1 \times \boldsymbol{\widehat{u}}_d}{\sqrt{1-(\boldsymbol{\widehat{u}}_1 \cdot \boldsymbol{\widehat{u}}_d)^2}} \times \boldsymbol{\widehat{u}}_1  .
\end{equation}
The scalar product of $\boldsymbol{\widehat{e}}_x$ and the normalized projection of $\boldsymbol{\widehat{u}}_2$ onto the $x$-$y$-plane is used to calculate $\phi_2$. The projection is achieved by discarding the $z$-component and normalizing the resulting vector. Normalizing the projection corresponds to a division by $\sin(\theta_2)= \sqrt{1-(\boldsymbol{\widehat{u}}_1 \cdot \boldsymbol{\widehat{u}}_2)^2} $.
As $\boldsymbol{\widehat{e}}_x$ is orthogonal to the $z$-axis, the $z$-component of $\boldsymbol{\widehat{u}}_2$ does not need to be discarded and the normalization is sufficient. The angle $\phi_2$ is equal to the angle between the projections of $\boldsymbol{\widehat{u}}_2$ into the $x$-$y$-plane and the $x$-axis:
\begin{equation}
  \phi_2
  =\arccos\!\bigg(\frac{\boldsymbol{\widehat{u}}_2}{\sqrt{1-(\boldsymbol{\widehat{u}}_1 \cdot \boldsymbol{\widehat{u}}_2)^2}} \cdot \boldsymbol{\widehat{e}}_x\bigg).
\end{equation}
There are coordinate singularities at $\sin(\theta_1)=0$ and at $\sin(\theta_2)=0$, i.e., for $\theta_1\in \{0,\pi\}$ and $\theta_2\in \{0,\pi\}$. The singularity for $\theta_2 \in \{0,\pi\}$ (i.e., at the poles) is typical for spherical coordinates. In contrast, the singularity for $\theta_1 \in \{0,\pi\}$ (i.e., when $\boldsymbol{\widehat{u}}_1 \parallel \boldsymbol{\widehat{u}}_d$) occurs due to the choice of $x$- and $y$-axis, which makes $\phi_2$ ambiguous in this case.
\\
The pair-distribution function possesses several angular symmetries, in particular
\begin{subequations}
 \label{eq:all_symmetries}
\begin{align}
g(r, \theta_1, \theta_2, \phi_2) &= g(r, \theta_1, \theta_2, - \phi_2) \label{eq:symm1}  , \\
g(r, \theta_1, \theta_2, \phi_2) &= g(r,- \theta_1, \theta_2, \pi- \phi_2) \label{eq:symm2} ,  \\
g(r, \theta_1, \theta_2, \phi_2) &= g(r, \theta_1,- \theta_2,\pi -\phi_2 )\label{eq:symm3} \, .
\end{align}
\end{subequations}
By definition, the angles $\theta_1$ and $\theta_2$ are limited to the interval $[0,\pi]$. With the symmetries \eqref{eq:symm1}-\eqref{eq:symm3}, however, we can extend their definition to all angles in the interval $[0,2\pi[$, making the pair-distribution function 2$ \pi$-periodic in all angular parameters. This extension allows us to obtain a Fourier transformation of the pair-distribution function later. For simplicity, we always refer to the periodically extended version of the pair-distribution function in the following.
\\
We measured the angles with a resolution of $2^\circ$ resulting in $N_\mathrm{b}=180$ bins for each angular parameter for the interval $[0,2\pi[$. The bin size of the distance parameter $r$ was chosen in a non-uniform way.
Values smaller than $1.2 \sigma$ were sampled with a bin size of $0.005\sigma$, while for values smaller than $3\sigma$ a bin size of $0.02\sigma$ and for $3\sigma < r < 10\sigma$ a bin size of $0.05\sigma$ was chosen.
We did not sample for any correlations beyond a distance of $r=10\sigma$.
\section{\label{sec:results}RESULTS AND DISCUSSION}
\begin{figure}[htb]
\centering
\includegraphics[]{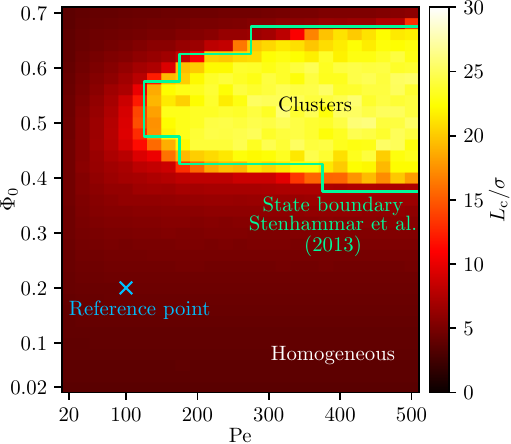}%
\caption{\label{fig2} State diagram for ABPs in three spatial dimensions. The characteristic length $L_\mathrm{c}$ described by Eq.\ \eqref{equ:char_length} quantifies density inhomogeneities which are characteristic for a clustered state thus allowing to identify regions of qualitatively different collective behavior.
For comparison, the state boundary found by Stenhammar et al.\ \cite{StenhammarMAC2014} is shown as well.
The blue cross indicates the reference point at $\mathrm{Pe} = 100$ and $\Phi_0 = 0.2$ used in the following figures.
A file containing the raw data for $L_\mathrm{c}$ shown here is provided as Supplementary Material \cite{SI}.
}
\end{figure}

The parametrization used for $g(\mathrm{Pe}, \Phi_0;r, \theta_1, \theta_2, \phi_2)$ requires the system to be in a homogeneous state.
Thus, we investigated the characteristic length $L_\mathrm{c}$ for different values of $\mathrm{Pe}$ and $\Phi_0$ 
to determine the regions in parameter space where the homogeneous state is stable.
The results are shown in Fig.\ \ref{fig2}.
It can be seen that the parameter space is split into two cohesive regions where either the homogeneous distribution is stable or phase separation occurs.
More precisely, the system remains homogeneous over time if either $\Phi_0<0.35$ or $\mathrm{Pe}<100$. Outside of this region, the system can exhibit phase separation. This becomes more favorable for high \Pecns and is suppressed only for very high densities. Here, we have a notable difference to the two-dimensional case discussed in Ref.\ \cite{JeggleSW2020}. In the two-dimensional case, the packing density of ABPs in a cluster can exceed the packing density of a perfect hexagonal structure of circles -- the highest possible packing density for impenetrable spheres -- due to overlapping \cite{StenhammarMAC2014}. Therefore, phase separation occurs. In the three-dimensional case, however, the packing density of particles in a cluster is lower than the perfect packing density of an fcc grid, even though slight overlapping is possible \cite{StenhammarMAC2014}. Consequently, the packing density of a \ZT{cluster} is not higher than the packing density of its environment, and therefore there is no phase separation. This argument, of course, hangs on the way we have defined the order parameter and therefore on our characterization of what \ZT{phase separation} is. If we, as done here, use the characteristic length, then the system will not be understood as being in a phase separated state if particles are everywhere (which is the case in three dimensions at high densities) since it then is homogeneous. In two dimensions, on the other hand, the fact that the particle density is extremely high in the cluster region has the consequence that there are fewer particles in another region, such that there is a finite characteristic length. 

\begin{figure*}[htb]
\centering
\includegraphics[]{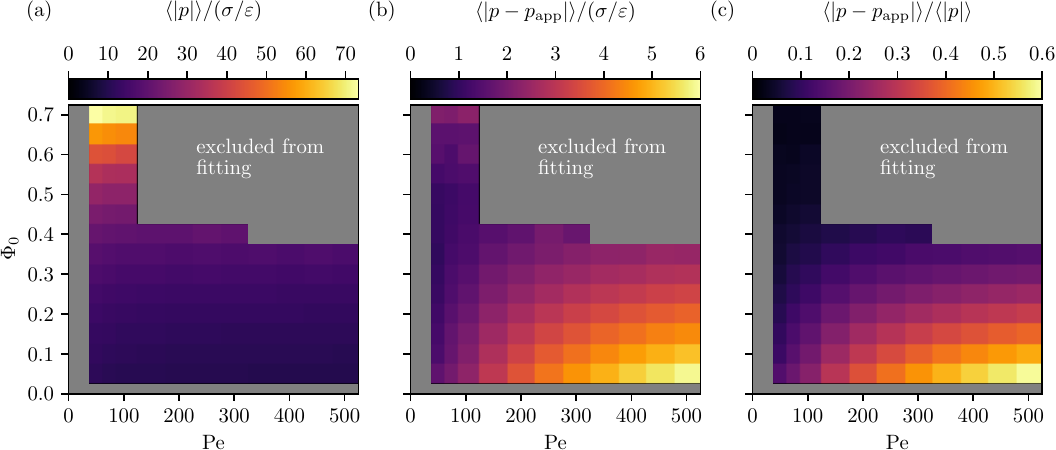}%
\caption{\label{fig3}
Areas excluded from fitting and results for the product function $p$ (see Section \ref{sec:results:analyticalapproximation}). (a) Mean absolute value (MAV) $\langle | p| \rangle/(\sigma/\epsilon)$ of the simulation data, (b) mean absolute error (MAE) $\langle |p - p_\mathrm{app}| \rangle/ (\sigma/\epsilon)$ between the simulation data and the analytical approximation, and (c) the relative error MAE/MAV.
The relative error is especially small for high packing densities and low \Pecnc and then grows with increasing \Pecn and shrinking packing density.
The grey areas are excluded from fitting, as phase separation was observed for these parameter combinations.}
\end{figure*}

Note that the state boundary found here is likely to be rather close to the spinodal. The reason is that, in the parameter region that is inside the binodal and outside the spinodal, both homogeneous states and clusters are (meta-)stable. In our simulations, we start with a homogeneous configuration and therefore end up in a homogeneous configuration in this region, but a simulation with an initial cluster could, in this region, have led to a final state with cluster formation.
\\
Our results confirm the state diagram by \citet{StenhammarMAC2014} and allow us to determine the regions where the symmetry assumptions introduced in Sec.\,\ref{subsec:parametrization} are valid (namely the regions where the particle distribution is homogeneous). The parameter combinations for which samples were taken to approximate $g$ analytically are shown in Fig.\ \ref{fig3}. %
\\
To check whether this state diagram contains finite size effects, we also investigated the state diagram of a system with twice the size in each dimension resulting in an eight times higher particle number.
The resulting state boundary widens up slightly, indicating phase separation for marginally smaller and higher densities. The critical \Pecn and density are not affected by the size of the system.
\subsection{\label{sec:results:pairdistributionfunction}Pair-distribution function}
\begin{figure*}[htb]
\centering
\includegraphics[]{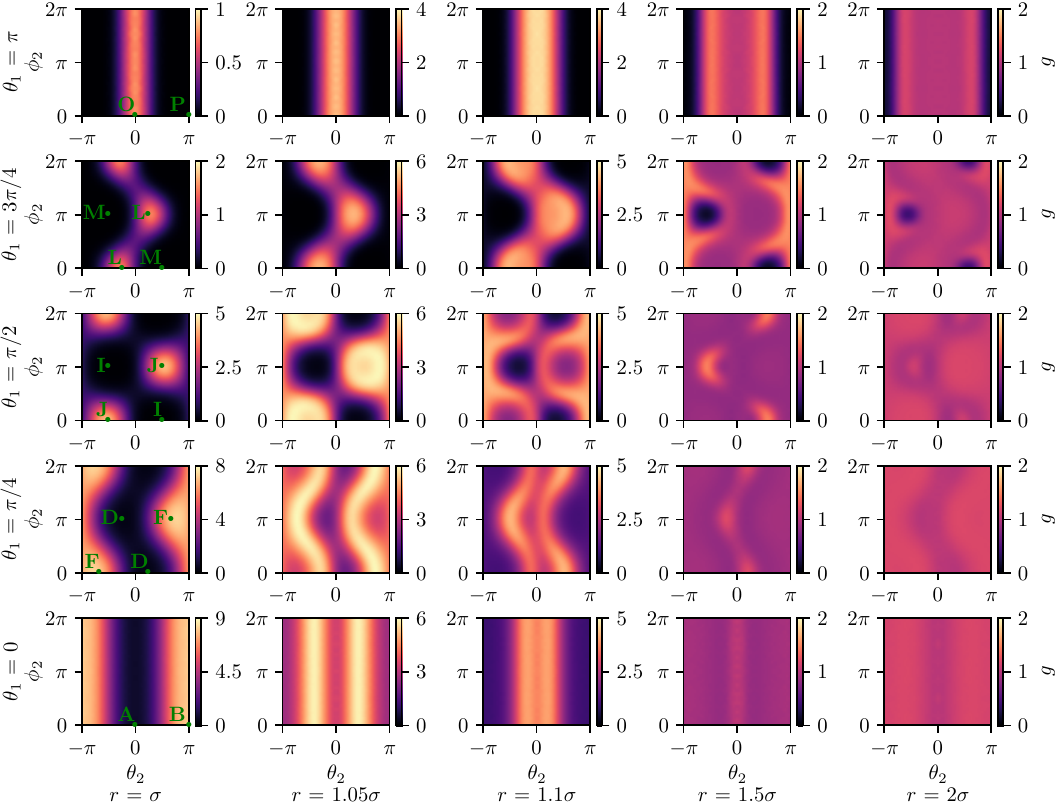}%
\caption{\label{fig4}Pair-distribution function $g(\mathrm{Pe}, \Phi_0;r , \theta_1, \theta_2, \phi_2)$ for distances $r\in \{\sigma, 1.05 \sigma, 1.1 \sigma, 1.5 \sigma, 2 \sigma\}$ and angles $\theta_1 \in \{0, \pi/4, \pi/2, 3\pi/4, \pi \}$ with fixed parameter values $\mathrm{Pe} = 100$ and $\Phi_0 = 0.2$.
This figure indicates the probability of finding a second particle with a prescribed position at different polar and azimuthal angles relative to a particle.
The configurations marked in green, which are shown in Fig.\ \ref{fig5}, represent extrema of this probability.}
\end{figure*}

\begin{figure*}[htb]
\centering
\includegraphics[]{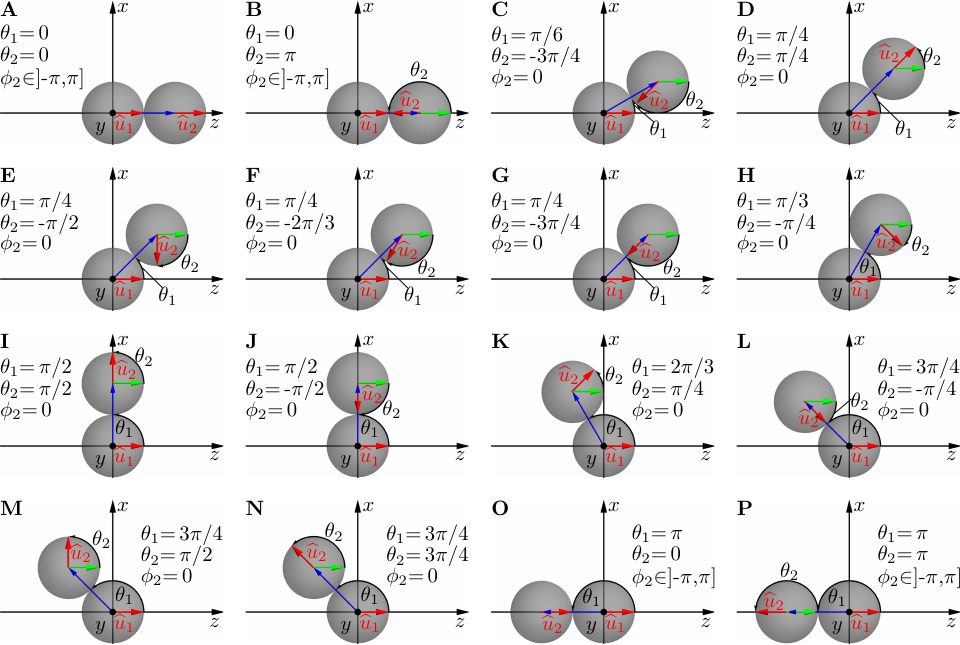}%
\caption{\label{fig5}Exemplary configurations of the particles that correspond to local extrema of the pair-distribution function $g$ at $r=\sigma$. These configurations are marked as green dots in Figs.\ \ref{fig4}, \ref{fig6}, and \ref{fig7}.}
\end{figure*}

\begin{figure*}[htb]
\centering
\includegraphics[]{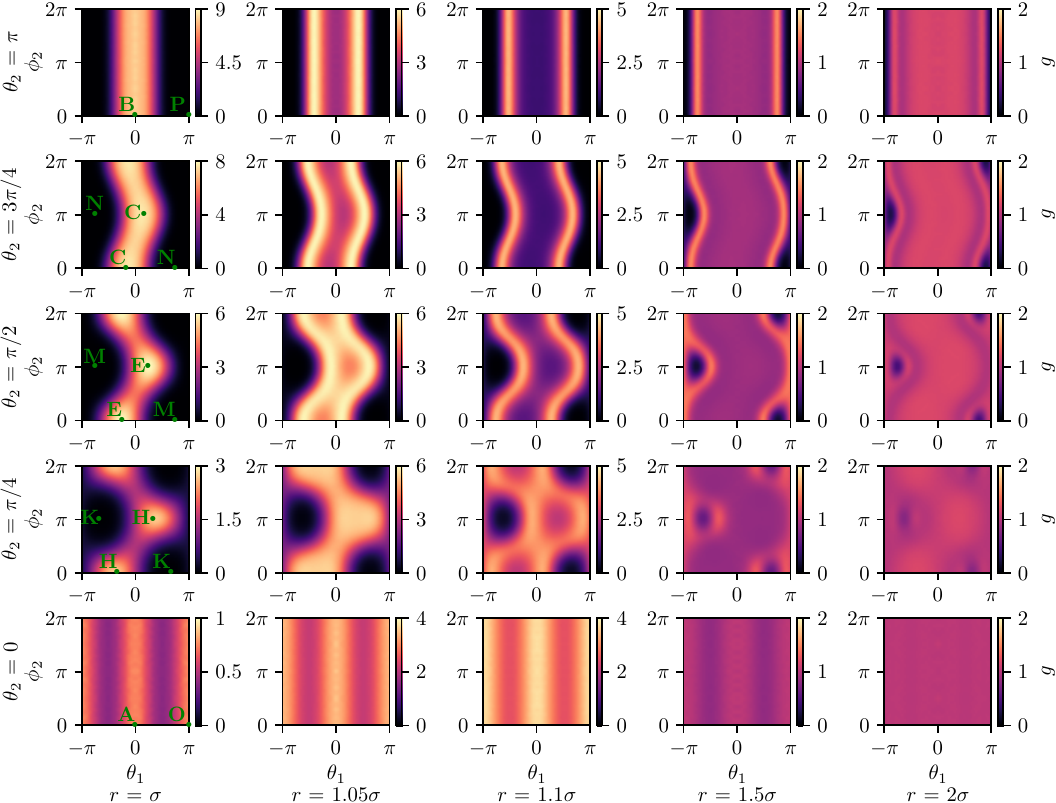}%
\caption{\label{fig6}Pair-distribution function $g(\mathrm{Pe}, \Phi_0;r , \theta_1 , \theta_2 , \phi_2)$ for distances $r\in \{\sigma, 1.05 \sigma, 1.1 \sigma, 1.5 \sigma, 2 \sigma\}$ and angles $\theta_2 \in \{0, \pi/4, \pi/2, 3\pi/4, \pi\}$ with fixed parameter values $\mathrm{Pe} = 100$ and $\Phi_0 = 0.2$.
This figure indicates the probability of finding a second particle with a prescribed polar angle at different positions and azimuthal angles relative to a particle.
The configurations marked in green, which are shown in Fig.\ \ref{fig5}, represent extrema of this probability.}
\end{figure*}

\begin{figure*}[htb]
\centering
\includegraphics[]{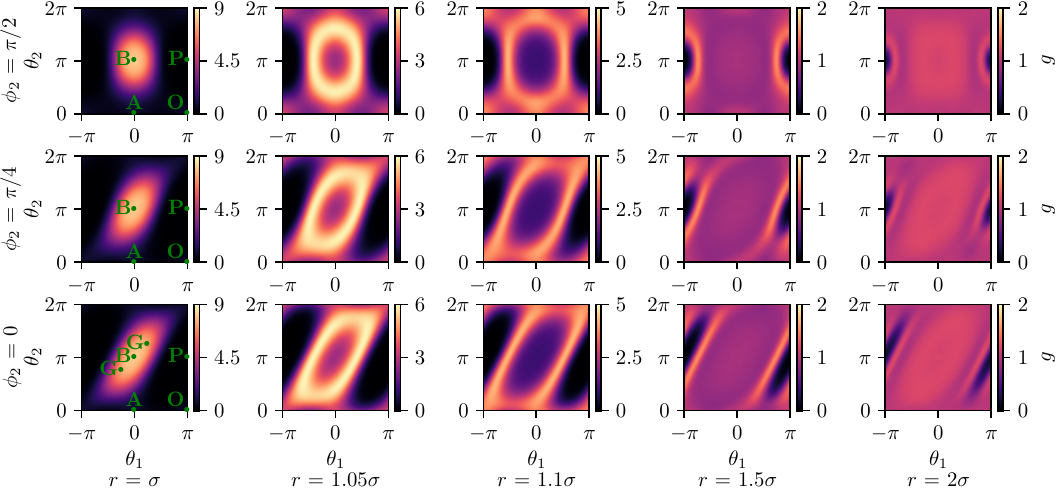}%
\caption{\label{fig7}Pair-distribution function $g(\mathrm{Pe}, \Phi_0;r, \theta_1 , \theta_2 , \phi_2)$ for distances $r\in \{\sigma, 1.05 \sigma, 1.1 \sigma, 1.5 \sigma, 2 \sigma \} $ and angles $\phi_2 \in \{0, \pi/4, \pi/2\} $ with fixed parameter values $\mathrm{Pe} = 100$ and $\Phi_0 = 0.2$.
This figure indicates the probability of finding a second particle with a prescribed azimuthal angle at different positions and polar angles relative to a particle.
The configurations marked in green, which are shown in Fig.\ \ref{fig5}, represent extrema of this probability. }
\end{figure*}

The pair-distribution function is shown for selected configurations in Figs.\ \ref{fig4}, \ref{fig6}, and \ref{fig7} for $\mathrm{Pe}=100$ and $\Phi_0=0.2$. These values have been chosen since they are deep in the homogeneous state (see Fig.\ \ref{fig2}).\\
Finding that the pair-distribution function can, with high accuracy, be represented by 15 Fourier modes, we use a low pass filter to cut off high frequencies in the angular dependence at the frequency $\omega_\mathrm{cut} = 15\cdot 2\pi =30\pi $ to minimize statistical errors. For large distances, $g$ goes to 1 since the probability of finding a particle at a position $\boldsymbol{r}_2$ is not influenced by the fact that another particle is at position $\boldsymbol{r}_1$ if $\boldsymbol{r}_1$ and $\boldsymbol{r}_2$ are very far apart.\\
We present $g$ for values of $r \in \{ \sigma, 1.05 \sigma, 1.1 \sigma, 1.5 \sigma, 2 \sigma \}$, which allows us to capture the most significant features of the pair-distribution function.
Figure\ \ref{fig4} shows the pair-distribution function for a selection of fixed distances $r$ and angles $\theta_1$. For each pair $(r,\theta_1)$, we plot $g$ as a function of $\theta_2$ and $\phi_2$. Therefore, each plot in Fig.\ \ref{fig4} corresponds to a fixed position of the second particle relative to the first particle and indicates the probability for every possible orientation of the second particle. Similarly, Fig.\ \ref{fig6} presents $g$ for fixed distances $r$ and polar angles $\theta_2$. For each pair $(r,\theta_2)$, $g$ is plotted as a function of $\theta_1$ and $\phi_2$. Finally, Fig.\ \ref{fig7} displays $g$ for fixed distances $r$ and azimuthal angles of the second particle $\phi_2$. For each pair $(r,\phi_2)$, $g$ is plotted as a function of $\theta_1$ and $\theta_2$. Certain configurations, which are highlighted in Figs.\ \ref{fig4}, \ref{fig6}, and \ref{fig7}, are visualized in Fig.\ \ref{fig5}.

In general, maxima of $g$ are found for configurations that are very stable or easy to reach. Similarly, minima of $g$ are found for configurations that are very unstable or impossible to reach. An example of a maximum of $g$ is the configuration \textbf{B} in Fig.\ \ref{fig5} with $\theta_1 = 0$, $\theta_2 = \pi$, and $\phi_2 \in ] - \pi, \pi ]$. In this case, the two particles are oriented towards each other and thus obstruct each other's motion until diffusion breaks up this configuration.
Therefore, this configuration is very stable.
In contrast, the configuration \textbf{P} in Fig.\ \ref{fig5} with $\theta_1 =\pi$, $\theta_2 = \pi$, and $\phi_2 \in ] - \pi, \pi ]$, i.e., two particles facing away from each other, coincides with a minimum of $g$. The reason is that (if we ignore thermal fluctuations) the only way to reach this configuration is that the particles move through each other (which is not possible for the interaction potential chosen here). 
As the repulsive interaction potential extends slightly further than $r=\sigma$, the pair-distribution function yields local maxima for $r=\sigma$ where the particles' propulsion force pushes the particles towards each other (e.g., configurations \textbf{B}, \textbf{F}, \textbf{J}, \textbf{L}, and \textbf{O} of Fig.\ \ref{fig5} marked in Fig.\ \ref{fig4}) and minima, where the interparticle force is minimized (e.g., configurations \textbf{A}, \textbf{D}, \textbf{I}, \textbf{M}, and \textbf{P} of Fig.\ \ref{fig5} marked in Fig.\ \ref{fig4}).
This line of reasoning also applies to the configurations in Figs.\ \ref{fig6} and \ref{fig7}. \\

For slightly larger distances such as $r = 1.05 \sigma$ and $r=1.1 \sigma$, we find that the maxima broaden and create small local minima in their center.
If the particles do not perfectly face each other, the interparticle force and propulsive force balance each other at a slightly larger separation. \\

The distribution function shows a change of structure for distances around $r=1.5 \sigma$.
As particles cannot pass through each other, but, due to the overdamped motion, also do not bounce back after a collision \cite{Loewen2020}, colliding particles often slide past each other. Therefore, configurations which result from particles moving past each other at a very small distance are more likely than configurations which result from particles moving past each other at a larger distance, as the latter happens only at random and not systematically due to interactions. 

In similar cases, a configuration resulting from particles sliding past each other (high probability) and a configuration that can essentially only be reached by particles passing through each other (low probability) are only separated by small offsets in the respective angles. Therefore, some configurations that can practically only emerge from particles passing through each other, such as the configurations \textbf{M}, \textbf{N}, and \textbf{K}, are surrounded by local maxima. 
\\ \\
If the azimuthal angle $\phi_2$ is zero, both orientation vectors $\boldsymbol{\widehat{u}}_1$ and $\boldsymbol{\widehat{u}}_2$ and the connecting vector $\boldsymbol{u}_d$ lie in the same plane, such that the configuration is quasi-two-dimensional. The pair-distribution function for this scenario is plotted in the bottom row of Fig.\ \ref{fig7}.
Note that the pair-distribution function is very similar to the pair-distribution function for a two-dimensional system obtained in Ref.\ \cite{JeggleSW2020}, which confirms our results.\\
We find that varying the \Pecn does not change the general structure of the pair-distribution function.
Increasing the \Pecn and thus reducing the temperature merely sharpens the features of $g$ in the form of taller and more narrow peaks of probability.
This makes sense considering that probability peaks in $g$ representing stable particle configurations are widened by rotational and translational diffusion, which diminish with decreasing temperatures. 
\\ 
In the case of low densities, in most cases only two particles interact with each other at a time.
If the density increases, the probability of interactions between three or more particles increases. If several particles are involved in a collision, the resulting interactions that determine the pair-distribution function become more complex and the structure of $g$ therefore becomes less sharp. Consequently, the effect of increasing the density is to broaden the maxima and minima of $g$. 
\subsection{\label{sec:results:analyticalapproximation}Analytical approximation of the function $-gU^\prime$}
One of the main reasons why the pair-distribution function is important is that it is required for deriving field theories \cite{StenhammarTAMC2013,WittkowskiSC2017,BialkeLS2013,SpeckMBL2015,teVrugtBW2022}. To see why, note that the dynamics of $\varrho$ for a system of ABPs in three dimensions is given by \cite{BickmannW2020b}
\begin{equation}
\begin{split}%
\dot{\varrho} &= (D_\mathrm{t}\Nabla_{\boldsymbol{r}_1}^2+D_\mathrm{r}\boldsymbol{\mathcal{R}_1}^2- v_0\Nabla_{\boldsymbol{r}_1}\cdot\boldsymbol{\widehat{u}}_1) \varrho + \mathcal{I}_\mathrm{int} \label{eqn:NeededforV}%
\end{split}%
\end{equation}
with the interaction term
\begin{equation}
\begin{split}
\mathcal{I}_\mathrm{int} &= \frac{D_\mathrm{t}}{k_\mathrm{B} T}\Nabla_{\boldsymbol{r}}\cdot \bigg( \varrho(\boldsymbol{r}, \boldsymbol{\widehat{u}}, t)\int_{\mathbb{R}^3}\!\!\!\!\!\dif^3r_2\, U_2'(\norm{\boldsymbol{r}_1-\boldsymbol{r}_2})\\
&\quad\,\:\! \frac{\boldsymbol{r}_1-\boldsymbol{r}_2}{\norm{\boldsymbol{r}_1 - \boldsymbol{r}_1}} \int_{\mathbb{S}_2}\!\!\!\!\dif^{2}u_2\, g(\boldsymbol{r}_1, \boldsymbol{r}_2, \boldsymbol{\widehat{u}_1}, \boldsymbol{\widehat{u}}_2, t)
\varrho(\boldsymbol{r}_1, \boldsymbol{\widehat{u}}_2, t) \bigg).
\end{split}
\label{eqn:Iint}%
\end{equation}
Here $U^\prime (r)=-\dif U_2(r)/\dif r$ is the interparticle force and $\mathcal{R}_i=\boldsymbol{\widehat{u}_i}\otimes\Nabla_{\boldsymbol{\widehat{u}}_i}$ the rotational operator. If (an approximation for) $g$ is not known, we cannot (not even approximately) evaluate the integral in \cref{eqn:Iint}. Therefore, field-theoretical models such as \cref{eqn:NeededforV} require an approximate analytical expression for $g$ as an input. Once such an expression has been provided, the then closed dynamic equation for $\varrho$ can be approximated further. This is done in the interaction-expansion method \cite{WittkowskiSC2017, BickmannW2020a, BickmannW2020b,BickmannBJW,teVrugtFHHTW2022}, which is reviewed in Ref.\ \cite{teVrugtBW2022}.

Taking a closer look at \cref{eqn:Iint}, we can see that what we actually require is not $g$, but the product of $g$ and $U^\prime$. Therefore, we now develop an analytical representation \footnote{In principle, since the interaction force is known in a microscopic simulation, the result also allows to calculate the pair-distribution function in the region where the force does not vanish. However, since the fit minimizes the error for $p$ rather than for $g$, it is not guaranteed that the resulting analytical expression for $g$ is always accurate.} for the product function
\begin{equation}
  p(\mathrm{Pe}, \Phi_0;r, \theta_1, \theta_2, \phi_2) = -U^\prime (r) \, g(\mathrm{Pe}, \Phi_0;r, \theta_1, \theta_2, \phi_2) \, ,
\end{equation}
which can be interpreted as a \ZT{pair-interaction-force distribution}.
In contrast to the pair-distribution function $g$, the product function $p$ is 
nonzero only for $0.9\sigma < r < 2^{1/6} \sigma$ \footnote{It is negligible for $r<0.9\sigma$ and exactly zero for $r>2^{1/6}\sigma$.}. The reason for this is that $g$ vanishes for $r$ less than approximately $\sigma$ due to the strong repulsion and $U^\prime$ vanishes for $r> 2^{1/6} \sigma$ due to the cutoff in the interaction force. 
Thus, $p$ only needs to be fitted in a narrow interval of $r$.
\\ \\
The function $p$ depends on the three angles $\theta_1$, $\theta_2$, and $\phi_2$, the distance $r$, the \Pecn Pe, and the packing density $\Phi_0$. First, we perform the real Fourier expansion %
\begin{align} \begin{split}
  &p(\mathrm{Pe}, \Phi_0;r, \theta_1, \theta_2, \phi_2)  = \\ &\sum_{h,j,k=0}^\infty \  \sum_{l,m,n=1}^{2} \,
  a_{h,j,k}^{l,m,n}(\mathrm{Pe}, \Phi_0;r) \, w_l(h\theta_1) w_m(j \theta_2 ) w_n(k \phi_2)
  \end{split}
\end{align}
with
\begin{align}
  w_1(x) &= \cos(x), \\ 
  w_2(x) &= \sin(x),
\end{align}
and
\begin{align}
 \begin{split}
&a_{h,j,k}^{l,m,n}(\mathrm{Pe}, \Phi_0;r) = \\ 
&\frac{1}{\pi^3} \int_0^{2\pi} \!\! \dif \theta_1 
\int_0^{2\pi} \!\! \dif \theta_2 
\int_0^{2\pi} \!\! \dif \phi_2 \ p(\mathrm{Pe}, \Phi_0;r, \theta_1, \theta_2, \phi_2) \\
  &w_l(h\theta_1) w_m(j \theta_2 ) w_n(k \phi_2) \, 2^{-(\delta_{h,0}+\delta_{j,0}+\delta_{k,0})}  \, .
  \end{split}
\end{align}
In our case, $p$ is not continuous but discrete since we use histograms for the data evaluation. This has to be accounted for in the definition of the coefficients via
\begin{align}
 \begin{split}
&a_{h,j,k}^{l,m,n}(\mathrm{Pe}, \Phi_0;r) = \\ &\frac{1}{\pi^3}
\sum_{o_1, o_2, o_3=0}^{N_\mathrm{b}-1}
\bigg( \frac{2\pi}{N_\mathrm{b}} \bigg)^3
 p\Big(\mathrm{Pe}, \Phi_0;r, 
\frac{N_\mathrm{b}}{2 \pi}  o_1,
\frac{N_\mathrm{b}}{2 \pi}  o_2,
\frac{N_\mathrm{b}}{2 \pi}  o_3\Big) \\
&w_l\Big(h\frac{N_\mathrm{b}}{2 \pi}  o_1\Big) 
 w_m\Big(j\frac{N_\mathrm{b}}{2 \pi}  o_2 \Big)
 w_n\Big(k\frac{N_\mathrm{b}}{2 \pi}  o_3\Big) \,
 2^{-(\delta_{h,0}+\delta_{j,0}+\delta_{k,0})}\, 
  \end{split}
\end{align}
with the number of bins $N_\mathrm{b}$. Using the symmetries of $g$ shown in Eqs.\,\eqref{eq:all_symmetries}, one can show that many coefficients vanish.
We find that the Fourier modes up to second order are sufficient for reproducing the structure of the product function reasonably well (similar to Ref.\ \cite{JeggleSW2020}) \footnote{The Fourier expansion truncated at the fifteenth order in Section \ref{sec:results:pairdistributionfunction} had the purpose of removing statistical errors. Here, the purpose of the Fourier expansion is to get tractable analytical expressions, which is why we truncate it already at second order.}. Figure \ref{fig3} shows that this truncation results only in a small error.
\quad This results in the approximation

\begin{align}
&p(\mathrm{Pe}, \Phi_0;r,\theta_1, \theta_2, \phi_2) \approx \notag\\ 
  & \sum_{h,j=0}^{2}  \  \sum_{k \in {0,2}}
  [\alpha_{h,j,k}(\mathrm{Pe}, \Phi_0;r) \cos ( h \theta_1) \cos( j \theta_2)\cos ( k \phi_2)] \notag\\
  &+ \sum_{h,j=1}^{2}  [\beta_{h,j,1}(\mathrm{Pe}, \Phi_0;r) \sin ( h \theta_1) \sin ( j \theta_2)\cos (  \phi_2) ]
\label{equ:p_approx_alpha_beta}
\end{align}
with $\alpha_{h,j,k} = a_{h,j,k}^{1,1,1}$ and $\beta_{h,j,k} = a_{h,j,k}^{2,2,1}$. We thus have 22 coefficients in total, each depending on $r$, Pe, and $\Phi_0$.
To fit the dependence of the coefficients on $r$, the product of an exponentially modified Gaussian distribution (EMG function) and a linear factor $(\sqrt[6]{2}-r)$ enforcing the cutoff was found to be useful.
The EMG function reads
\begin{align}
\begin{split}
  \mathrm{EMG}(r; \mu, \omega, \lambda) &= \frac{\lambda}{2} \exp  \! \Big( \frac{\lambda}{2} ( \lambda \omega^2  -2(r-\mu)) \Big)  \\
&\quad \: \:  \mathrm{erfc}\Big(\frac{ \lambda \omega^2 - (r-\mu)}{\sqrt{2} \omega}\Big),
\end{split}
\end{align}
where $\mu$ is the mean value, $\omega$ the standard deviation, $\lambda$ the rate of the exponential component which controls the skewness of the distribution, and $\mathrm{erfc}$ the complementary error function.\\
We found that all coefficients can be fitted by a product of the EMG function, the linear cutoff term, and a polynomial of a degree less than four.
\begin{figure*}[htb]
\centering
\includegraphics[]{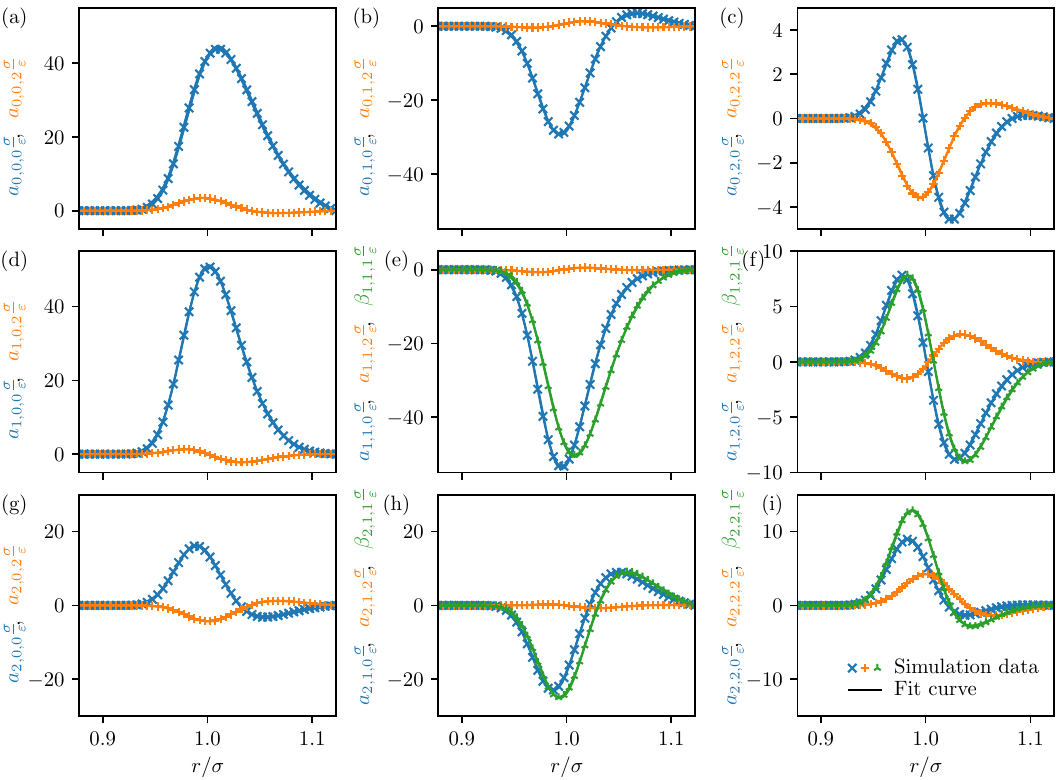}%
\caption{\label{fig8}Fourier coefficients $\alpha_{h,j,k}(\mathrm{Pe}, \Phi_0;r)$ and $\beta_{h,j,1}(\mathrm{Pe}, \Phi_0;r)$ for the simulation data (symbols) and the corresponding fitted functions $f_0,f_1,f_2,f_3$, and $f_4$ (solid lines) at the reference parameters $\mathrm{Pe}=100$ and $ \Phi_0=0.2$. The coefficients $\alpha_{h,j,k}$ and $\beta_{h,j,1}$ are plotted in the $h$-th row and the $j$-th column. Coefficients where the last index is zero 
are plotted with blue crosses, while green triangular symbols are used for coefficients where the last index is one, and orange plus symbols are used if the last index is two. Since $\beta_{h,j,k}=0$ for $h=0$ or $j=0$, there are only two rather than three curves in (a), (d), and (g). 
}
\end{figure*}
Overall, the functions used to fit the coefficients $\alpha_{h,j,k}$ and $\beta_{h,j,1}$ are
\begin{align}
  &f_0 (r;a,\mu, \omega, \lambda) = a \,  \text{EMG}(r; \mu, \omega, \lambda)  (\sqrt[6]{2}-r), \label{equ:fitfunction_mit_omega} \\
  &f_1 (r;a,\mu, \omega, \lambda,b)  =   f_0 (r;a,\mu, \omega, \lambda)  (b-r), \\
  &f_2 (r;a,\mu, \omega, \lambda,b,c)  =   f_1 (r;a,\mu, \omega, \lambda,b)    (c-r), \\
  &f_3 (r;a,\mu, \omega, \lambda,b,c)  =   f_0 (r;a,\mu, \omega, \lambda)  (r^2+br+c), \\
  &f_4 (r;a,\mu, \omega, \lambda,b,c,d)  =   f_3 (r;a,\mu, \omega, \lambda,b,c)  (d-r),
\end{align}
where $a$ is a scaling factor of the EMG function and $b$, $c$, and $d$ are additional fit parameters of the polynomial functions.
While $f_2$ is a special case of $f_3$ with the roots of the polynomial factor being purely real, we achieved higher numerical stability by fitting with $f_2$.
The coefficients and their corresponding fits for $\mathrm{Pe}=100$ and $\Phi_0=0.2$ are shown in Fig.\ \ref{fig8}.
We chose the fit function for each coefficient based on the number of zero-crossings and thus the required degree of the polynomial factor.
As each coefficient also depends on the \Pecn and the packing density,
the function to fit each coefficient is chosen according to the maximum number of zero-crossings observed for any value of Pe and $\Phi_0$.
If the number of zero-crossings changes with Pe or $\Phi_0$, the roots of the polynomial term can move out of the interval, where $p$ is nonzero, causing numerical instability. These cases are treated separately with adjusted starting values for the fit to obtain reasonably smooth curves for the coefficients in Pe-$\Phi_0$ parameter space.
\\
The described fitting procedure allows us to describe four of the six parameter dependencies of $p(\mathrm{Pe}, \Phi_0;r,\theta_1, \theta_2, \phi_2)$ analytically and only the dependencies of the fit parameters on Pe and $\Phi_0$ remain unknown.
In order to get a purely analytical expression for $p$, this dependence also needs to be interpolated for the region of parameter space shown in Fig.\ \ref{fig3}. For this we use the empirically motivated fit function
\begin{equation}
h(\mathrm{Pe}, \Phi_0) = \sum_{m=-2}^{2} \sum_{n=0}^{3} \mathrm{Pe}^{\frac{m}{2}} \Phi_0^n \urep_{m,n}
\end{equation}
with the fit parameters $\urep_{m,n}$. The results of these fits are given in Appendix \ref{sec:fitparameters}. \\ \\

To estimate the quality of our approximations, we calculate the deviation of our analytical approximations $p_\mathrm{app}$ from the numerical result for $p$ measured directly from simulation. We quantify the difference via the mean absolute error (MAE)
\begin{align}\begin{split}
  &\mathrm{MAE}(p_\mathrm{app},p) =\langle |p-p_\mathrm{app}| \rangle =  \\ &\
   \frac{\int_{r_{\mathrm{min}}}^{r_{\mathrm{max}}} \! \dif r \int_0^{\pi}\!\! \dif \theta_1 \int_0^{\pi} \!\! \dif \theta_2 \int_0^{2 \pi}\!\! \dif \phi_2   \sin (\theta_1) \sin(\theta_2) |p_\mathrm{app}-p| }
  {\int_{r_{\mathrm{min}}}^{r_{\mathrm{max}}} \! \dif r \int_0^{\pi} 
 \!\! \dif \theta_1 \int_0^{\pi}\!\! \dif \theta_2 \int_0^{2 \pi} \!\! \dif \phi_2   \sin (\theta_1) \sin(\theta_2)} \, .
\end{split}\end{align}
In this case, $r_\mathrm{max} $ equals $ \sqrt[6]{2}\,  \sigma$ as the interaction potential is zero for higher values of $r$ and $r_\mathrm{min} $ equals $ 0.8 \, \sigma$ as no two particles with a smaller distance were found in the simulations. For the relative error, we calculate the ratio between the MAE and the mean absolute value (MAV) of $p$
\begin{align}\begin{split}
  &\mathrm{MAV}(p) = \langle |p| \rangle   \\ &=
   \frac{\int_{r_{\mathrm{min}}}^{r_{\mathrm{max}}} \! \dif r \int_0^{\pi} \!\! \dif \theta_1 \int_0^{\pi} \!\! \dif \theta_2 \int_0^{2 \pi} \!\! \dif \phi_2   \sin (\theta_1) \sin(\theta_2) |p| }
  {\int_{r_{\mathrm{min}}}^{r_{\mathrm{max}}} \! \dif r \int_0^{\pi} \!\! \dif \theta_1 \int_0^{\pi}\!\! \dif \theta_2 \int_0^{2 \pi} \!\! \dif \phi_2   \sin (\theta_1) \sin(\theta_2)} \, .
\end{split}\end{align}
In Fig.\ \ref{fig3}, we show the MAE, the MAV, and the relative error MAE/MAV. The relative error varies between 2 and 55 percent with high errors occurring only for very small packing densities and very high \Pecnsp
We find that the errors are predominantly introduced by the frequency cutoff approximation and not by the two fitting steps.
The reason for the high errors observed for low densities and high \Pecn are the high frequency modes of $g$ in this regime that result from the steeper slope of $g$ (see Section \ref{sec:results:pairdistributionfunction}).
\section{\label{sec:conclusions}CONCLUSIONS}
In this work, we have obtained the state diagram and the full pair-distribution function of ABPs in three spatial dimensions using Brownian dynamics simulations. Our results confirm and improve state diagrams obtained in previous works \cite{StenhammarMAC2014,SiebertLSV2017,WysockiWG2014}. Note that the state boundary found in this work corresponds to the spinodal rather than to the binodal. Furthermore, the fully orientation-resolved pair-distribution function for homogeneous particle distributions has been extracted from the simulations for a wide range of \Pecns and packing densities. If our result is restricted to a two-dimensional plane, it agrees with the form obtained in Ref.\ \cite{JeggleSW2020} for a two-dimensional system. Exploiting translational, rotational, and temporal invariances, the pair-distribution function can be parametrized using only six parameters. An intuitive explanation for the form of the pair-distribution function has been provided. In addition, we found an analytical expression for the product of the pair-distribution function and the derivative of the interaction potential that provides an excellent fit to the simulation data. \\

Our work extends the results by \citet{JeggleSW2020} by adding a third spatial dimension, the results by \citet{SchwarzendahlM2018} by providing the full angular dependence of the pair-distribution function, and the results by \citet{DhontPB2021motility} by considering also the case of high densities. The consistency of our results with previous work is demonstrated by the agreement with Ref.\ \cite{JeggleSW2020} for two-dimensional cross sections. However, the different form of the state boundary for MIPS shows the importance of considering also the three-dimensional case in full detail. Our results provide interesting insights into the collective dynamics of ABPs in three spatial dimensions and can be exploited in the derivation of active field theories and for obtaining microscopic predictions for state boundaries in active systems \cite{WittkowskiSC2017, BickmannW2020a,BickmannW2020b, BialkeLS2013,SpeckMBL2015,BickmannW2020b,BickmannBJW,teVrugtBW2022}. 
In particular, our results have already been used in Ref. \cite{BickmannW2020b} for the derivation of a predictive field theory. Possible extensions, for which our results provide a useful starting point, are the investigation of  mixtures of active and passive particles and of particles with more complex shapes.
\section*{Supplementary Material}
The Supplementary Material \cite{SI} contains a spreadsheet with the values of the fit parameters (as shown in Appendix \ref{sec:fitparameters}) that are needed to recreate the analytical representation of the product function, a Python script \texttt{abp.spherical3d.pairdistribution} that recreates the approximation of the product function $p$ using the values of the fit parameters, and the Python scripts and raw data needed to recreate Figs.\ \ref{fig1}--\ref{fig8}. 
\section*{CONFLICTS OF INTEREST}
There are no conflicts of interests to declare.
\section*{ACKNOWLEDGEMENTS}
R.W.\ is funded by the Deutsche Forschungsgemeinschaft (DFG, German Research Foundation) -- Project-ID 433682494 -- SFB 1459.  
The simulations for this work were performed on the computer cluster PALMA II of the University of M\"unster.

\FloatBarrier
\twocolumngrid

\bibliography{refs}

\FloatBarrier
\onecolumngrid
\begin{appendix}

\section{\label{sec:fitparameters}Fit parameters}
In the following tables, we provide the optimal fit parameters to fit the
function $h(\mathrm{Pe}, \Phi_0) = \sum_{m=-2}^{2} \sum_{n=0}^{3} \mathrm{Pe}^{\frac{m}{2}} \Phi_0^n \urep_{m,n}$ to each parameter
used to fit the Fourier coefficients $\alpha$ and $\beta$. Reference \cite{SI} contains these parameters as .csv-data set for easier use and a Python program with a function that reads the data set and returns the value of $g(r, \boldsymbol{\widehat{u}}_d,\boldsymbol{\widehat{u}}_1,\boldsymbol{\widehat{u}}_2)$ as well as $g(r, \theta_1, \theta_2, \phi_2)$.
\begin{table*}[!htbp]
  \centering
\begin{tabularx}{\linewidth}{|c|c|Y|Y|Y|Y|Y|Y|Y|}
    \cline{3-9}
    \multicolumn{2}{c|}{} & $\urep_{-2,0}$  & $\urep_{-2, 1}$ & $\urep_{-2, 2}$ & $\urep_{ -2, 3}$ & $\urep_{-1 ,0 }$ & $\urep_{-1 ,1 }$ &$\urep_{-1 ,2 }$  \\
    \multicolumn{2}{c|}{} & $\urep_{ -1, 3}$& $\urep_{ 0, 0}$ & $\urep_{ 0,1 }$ & $\urep_{0 ,2 }$ & $\urep_{0 , 3}$ & $\urep_{1 ,0 }$ &$\urep_{1 ,1 }$  \\
    \multicolumn{2}{c|}{} & $\urep_{ 1,2 }$  & $\urep_{1 ,3 }$ & $\urep_{ 2,0 }$ & $\urep_{ 2, 1}$ & $\urep_{ 2,2 }$ & $\urep_{2 ,3 }$ & \\
\hline
& & & & & & & & \\[-1.01em]
$\alpha_{0,0,0}$ &   & $3.851\cdot 10^{3}$   & $-1.123\cdot 10^{3}$   & $1.455\cdot 10^{2}$ & $-3.513$   & $5.588\cdot 10^{-2}$   & $-6.308\cdot 10^{4}$& $2.123\cdot 10^{4}$ \\[-0.25em] \cline{3-9} & & & & & & & & \\[-1.01em]
 &$a$& $-2.379\cdot 10^{3}$   & $1.222\cdot 10^{2}$   & $-2.224$ & $2.768\cdot 10^{5}$   & $-9.911\cdot 10^{4}$   & $1.235\cdot 10^{4}$& $-6.949\cdot 10^{2}$ \\[-0.25em] \cline{3-9} & & & & & & & &  \\
\hline
\end{tabularx}
\caption{\label{tab0}  Fit coefficients of the function $h(\mathrm{Pe}, \Phi_0)$ used to fit the variables of the Fourier coefficients $\alpha_{0,0,0}$, $\alpha_{0,0,2}$, and $\alpha_{0,1,0}$ of the function $p$.}

\end{table*}

\begin{table*}[!htbp]
  \centering
  \begin{tabularx}{\linewidth}{|c|c|Y|Y|Y|Y|Y|Y|Y|}
    \cline{3-9}
    \multicolumn{2}{c|}{} & $\urep_{-2,0}$  & $\urep_{-2, 1}$ & $\urep_{-2, 2}$ & $\urep_{ -2, 3}$ & $\urep_{-1 ,0 }$ & $\urep_{-1 ,1 }$ &$\urep_{-1 ,2 }$  \\
    \multicolumn{2}{c|}{} & $\urep_{ -1, 3}$& $\urep_{ 0, 0}$ & $\urep_{ 0,1 }$ & $\urep_{0 ,2 }$ & $\urep_{0 , 3}$ & $\urep_{1 ,0 }$ &$\urep_{1 ,1 }$  \\
    \multicolumn{2}{c|}{} & $\urep_{ 1,2 }$  & $\urep_{1 ,3 }$ & $\urep_{ 2,0 }$ & $\urep_{ 2, 1}$ & $\urep_{ 2,2 }$ & $\urep_{2 ,3 }$ & \\
\hline
& & & & & & & & \\[-1.01em]
$\alpha_{0,1,2}$ &   & $3.634\cdot 10^{5}$   & $-1.563\cdot 10^{5}$   & $2.507\cdot 10^{4}$ & $-1.702\cdot 10^{3}$   & $2.754\cdot 10^{1}$   & $-4.355\cdot 10^{6}$& $1.626\cdot 10^{6}$ \\[-0.25em] \cline{3-9} & & & & & & & & \\[-1.01em]
 &$a$& $-2.133\cdot 10^{5}$   & $1.126\cdot 10^{4}$   & $-1.742\cdot 10^{2}$ & $1.759\cdot 10^{7}$   & $-6.411\cdot 10^{6}$   & $8.051\cdot 10^{5}$& $-3.937\cdot 10^{4}$ \\[-0.25em] \cline{3-9} & & & & & & & &  \\
\hline
\end{tabularx}
\caption{\label{tab1}  Table of fit coefficients analog to Tab.\ \ref{tab0}, but for the Fourier coefficients $\alpha_{0,1,2}$, $\alpha_{0,2,0}$, and $\alpha_{0,2,2}$.}

\end{table*}

\begin{table*}[!htbp]
  \centering
  \begin{tabularx}{\linewidth}{|c|c|Y|Y|Y|Y|Y|Y|Y|}
    \cline{3-9}
    \multicolumn{2}{c|}{} & $\urep_{-2,0}$  & $\urep_{-2, 1}$ & $\urep_{-2, 2}$ & $\urep_{ -2, 3}$ & $\urep_{-1 ,0 }$ & $\urep_{-1 ,1 }$ &$\urep_{-1 ,2 }$  \\
    \multicolumn{2}{c|}{} & $\urep_{ -1, 3}$& $\urep_{ 0, 0}$ & $\urep_{ 0,1 }$ & $\urep_{0 ,2 }$ & $\urep_{0 , 3}$ & $\urep_{1 ,0 }$ &$\urep_{1 ,1 }$  \\
    \multicolumn{2}{c|}{} & $\urep_{ 1,2 }$  & $\urep_{1 ,3 }$ & $\urep_{ 2,0 }$ & $\urep_{ 2, 1}$ & $\urep_{ 2,2 }$ & $\urep_{2 ,3 }$ & \\
\hline
& & & & & & & & \\[-1.01em]
$\alpha_{1,0,0}$ &   & $3.223\cdot 10^{4}$   & $-2.288\cdot 10^{3}$   & $6.567\cdot 10^{2}$ & $1.037\cdot 10^{1}$   & $-3.946\cdot 10^{-1}$   & $1.926\cdot 10^{5}$& $-7.301\cdot 10^{4}$ \\[-0.25em] \cline{3-9} & & & & & & & & \\[-1.01em]
 &$a$& $4.908\cdot 10^{3}$   & $-5.720\cdot 10^{1}$   & $2.264\cdot 10^{-2}$ & $-6.046\cdot 10^{5}$   & $9.449\cdot 10^{4}$   & $9.811\cdot 10^{3}$& $-1.600\cdot 10^{3}$ \\[-0.25em] \cline{3-9} & & & & & & & &  \\
\hline
\end{tabularx}
\caption{\label{tab2}  Table of fit coefficients analog to Tab.\ \ref{tab0}, but for the Fourier coefficients $\alpha_{1,0,0}$, $\alpha_{1,0,2}$, and $\alpha_{1,1,0}$.}

\end{table*}

\begin{table*}[!htbp]
  \centering
  \begin{tabularx}{\linewidth}{|c|c|Y|Y|Y|Y|Y|Y|Y|}
    \cline{3-9}
    \multicolumn{2}{c|}{} & $\urep_{-2,0}$  & $\urep_{-2, 1}$ & $\urep_{-2, 2}$ & $\urep_{ -2, 3}$ & $\urep_{-1 ,0 }$ & $\urep_{-1 ,1 }$ &$\urep_{-1 ,2 }$  \\
    \multicolumn{2}{c|}{} & $\urep_{ -1, 3}$& $\urep_{ 0, 0}$ & $\urep_{ 0,1 }$ & $\urep_{0 ,2 }$ & $\urep_{0 , 3}$ & $\urep_{1 ,0 }$ &$\urep_{1 ,1 }$  \\
    \multicolumn{2}{c|}{} & $\urep_{ 1,2 }$  & $\urep_{1 ,3 }$ & $\urep_{ 2,0 }$ & $\urep_{ 2, 1}$ & $\urep_{ 2,2 }$ & $\urep_{2 ,3 }$ & \\
\hline
& & & & & & & & \\[-1.01em]
$\alpha_{1,1,2}$ &   & $-9.954\cdot 10^{5}$   & $2.120\cdot 10^{5}$   & $1.642\cdot 10^{4}$ & $-3.131\cdot 10^{3}$   & $-2.341\cdot 10^{2}$   & $6.628\cdot 10^{7}$& $-3.079\cdot 10^{7}$ \\[-0.25em] \cline{3-9} & & & & & & & & \\[-1.01em]
 &$a$& $4.842\cdot 10^{6}$   & $-3.063\cdot 10^{5}$   & $7.906\cdot 10^{3}$ & $-2.906\cdot 10^{8}$   & $1.332\cdot 10^{8}$   & $-2.123\cdot 10^{7}$& $1.364\cdot 10^{6}$ \\[-0.25em] \cline{3-9} & & & & & & & &  \\
\hline
\end{tabularx}
\caption{\label{tab3}  Table of fit coefficients analog to Tab.\ \ref{tab0}, but for the Fourier coefficients $\alpha_{1,1,2}$, $\alpha_{1,2,0}$, and $\alpha_{1,2,2}$.}

\end{table*}

\begin{table*}[!htbp]
  \centering
  \begin{tabularx}{\linewidth}{|c|c|Y|Y|Y|Y|Y|Y|Y|}
    \cline{3-9}
    \multicolumn{2}{c|}{} & $\urep_{-2,0}$  & $\urep_{-2, 1}$ & $\urep_{-2, 2}$ & $\urep_{ -2, 3}$ & $\urep_{-1 ,0 }$ & $\urep_{-1 ,1 }$ &$\urep_{-1 ,2 }$  \\
    \multicolumn{2}{c|}{} & $\urep_{ -1, 3}$& $\urep_{ 0, 0}$ & $\urep_{ 0,1 }$ & $\urep_{0 ,2 }$ & $\urep_{0 , 3}$ & $\urep_{1 ,0 }$ &$\urep_{1 ,1 }$  \\
    \multicolumn{2}{c|}{} & $\urep_{ 1,2 }$  & $\urep_{1 ,3 }$ & $\urep_{ 2,0 }$ & $\urep_{ 2, 1}$ & $\urep_{ 2,2 }$ & $\urep_{2 ,3 }$ & \\
\hline
& & & & & & & & \\[-1.01em]
$\alpha_{2,0,0}$ &   & $2.232\cdot 10^{4}$   & $-8.683\cdot 10^{3}$   & $1.043\cdot 10^{3}$ & $1.523\cdot 10^{-1}$   & $1.598\cdot 10^{-1}$   & $-1.126\cdot 10^{5}$& $6.003\cdot 10^{4}$ \\[-0.25em] \cline{3-9} & & & & & & & & \\[-1.01em]
 &$a$& $-1.080\cdot 10^{4}$   & $6.921\cdot 10^{2}$   & $-1.601\cdot 10^{1}$ & $3.951\cdot 10^{5}$   & $-2.471\cdot 10^{5}$   & $4.932\cdot 10^{4}$& $-3.565\cdot 10^{3}$ \\[-0.25em] \cline{3-9} & & & & & & & &  \\
\hline
\end{tabularx}
\caption{\label{tab4}  Table of fit coefficients analog to Tab.\ \ref{tab0}, but for the Fourier coefficients $\alpha_{2,0,0}$, $\alpha_{2,0,2}$, and $\alpha_{2,1,0}$.}

\end{table*}

\begin{table*}[!htbp]
  \centering
  \begin{tabularx}{\linewidth}{|c|c|Y|Y|Y|Y|Y|Y|Y|}
    \cline{3-9}
    \multicolumn{2}{c|}{} & $\urep_{-2,0}$  & $\urep_{-2, 1}$ & $\urep_{-2, 2}$ & $\urep_{ -2, 3}$ & $\urep_{-1 ,0 }$ & $\urep_{-1 ,1 }$ &$\urep_{-1 ,2 }$  \\
    \multicolumn{2}{c|}{} & $\urep_{ -1, 3}$& $\urep_{ 0, 0}$ & $\urep_{ 0,1 }$ & $\urep_{0 ,2 }$ & $\urep_{0 , 3}$ & $\urep_{1 ,0 }$ &$\urep_{1 ,1 }$  \\
    \multicolumn{2}{c|}{} & $\urep_{ 1,2 }$  & $\urep_{1 ,3 }$ & $\urep_{ 2,0 }$ & $\urep_{ 2, 1}$ & $\urep_{ 2,2 }$ & $\urep_{2 ,3 }$ & \\
\hline
& & & & & & & & \\[-1.01em]
$\alpha_{2,1,2}$ &   & $3.221\cdot 10^{5}$   & $-8.918\cdot 10^{4}$   & $7.327\cdot 10^{3}$ & $-2.778\cdot 10^{2}$   & $2.294\cdot 10^{1}$   & $-6.844\cdot 10^{6}$& $2.531\cdot 10^{6}$ \\[-0.25em] \cline{3-9} & & & & & & & & \\[-1.01em]
 &$a$& $-3.415\cdot 10^{5}$   & $2.053\cdot 10^{4}$   & $-5.234\cdot 10^{2}$ & $2.563\cdot 10^{7}$   & $-1.001\cdot 10^{7}$   & $1.434\cdot 10^{6}$& $-9.006\cdot 10^{4}$ \\[-0.25em] \cline{3-9} & & & & & & & &  \\
\hline
\end{tabularx}
\caption{\label{tab5}  Table of fit coefficients analog to Tab.\ \ref{tab0}, but for the Fourier coefficients $\alpha_{2,1,2}$, $\alpha_{2,2,0}$, and $\alpha_{2,2,2}$.}

\end{table*}

\begin{table*}[!htbp]
  \centering
  \begin{tabularx}{\linewidth}{|c|c|Y|Y|Y|Y|Y|Y|Y|}
    \cline{3-9}
    \multicolumn{2}{c|}{} & $\urep_{-2,0}$  & $\urep_{-2, 1}$ & $\urep_{-2, 2}$ & $\urep_{ -2, 3}$ & $\urep_{-1 ,0 }$ & $\urep_{-1 ,1 }$ &$\urep_{-1 ,2 }$  \\
    \multicolumn{2}{c|}{} & $\urep_{ -1, 3}$& $\urep_{ 0, 0}$ & $\urep_{ 0,1 }$ & $\urep_{0 ,2 }$ & $\urep_{0 , 3}$ & $\urep_{1 ,0 }$ &$\urep_{1 ,1 }$  \\
    \multicolumn{2}{c|}{} & $\urep_{ 1,2 }$  & $\urep_{1 ,3 }$ & $\urep_{ 2,0 }$ & $\urep_{ 2, 1}$ & $\urep_{ 2,2 }$ & $\urep_{2 ,3 }$ & \\
\hline
& & & & & & & & \\[-1.01em]
$\beta_{1,1,1}$ &   & $1.819\cdot 10^{5}$   & $-8.367\cdot 10^{4}$   & $8.197\cdot 10^{3}$ & $-2.983\cdot 10^{2}$   & $-4.344$   & $-1.359\cdot 10^{6}$& $4.460\cdot 10^{5}$ \\[-0.25em] \cline{3-9} & & & & & & & & \\[-1.01em]
 &$a$& $-2.272\cdot 10^{4}$   & $-1.699\cdot 10^{3}$   & $1.455\cdot 10^{2}$ & $4.377\cdot 10^{5}$   & $4.344\cdot 10^{5}$   & $-2.058\cdot 10^{5}$& $2.346\cdot 10^{4}$ \\[-0.25em] \cline{3-9} & & & & & & & &  \\
\hline
\end{tabularx}
\caption{\label{tab6}  Table of fit coefficients analog to Tab.\ \ref{tab0}, but for the Fourier coefficients $\beta_{1,1,1}$ and $\beta_{1,2,1}$.}

\end{table*}

\begin{table*}[!htbp]
  \centering
  \begin{tabularx}{\linewidth}{|c|c|Y|Y|Y|Y|Y|Y|Y|}
    \cline{3-9}
    \multicolumn{2}{c|}{} & $\urep_{-2,0}$  & $\urep_{-2, 1}$ & $\urep_{-2, 2}$ & $\urep_{ -2, 3}$ & $\urep_{-1 ,0 }$ & $\urep_{-1 ,1 }$ &$\urep_{-1 ,2 }$  \\
    \multicolumn{2}{c|}{} & $\urep_{ -1, 3}$& $\urep_{ 0, 0}$ & $\urep_{ 0,1 }$ & $\urep_{0 ,2 }$ & $\urep_{0 , 3}$ & $\urep_{1 ,0 }$ &$\urep_{1 ,1 }$  \\
    \multicolumn{2}{c|}{} & $\urep_{ 1,2 }$  & $\urep_{1 ,3 }$ & $\urep_{ 2,0 }$ & $\urep_{ 2, 1}$ & $\urep_{ 2,2 }$ & $\urep_{2 ,3 }$ & \\
\hline
& & & & & & & & \\[-1.01em]
$\beta_{2,1,1}$ &   & $-2.553\cdot 10^{4}$   & $1.023\cdot 10^{4}$   & $-1.124\cdot 10^{3}$ & $-5.362\cdot 10^{1}$   & $8.654\cdot 10^{-1}$   & $-3.253\cdot 10^{5}$& $1.089\cdot 10^{5}$ \\[-0.25em] \cline{3-9} & & & & & & & & \\[-1.01em]
 &$a$& $-1.141\cdot 10^{4}$   & $4.098\cdot 10^{2}$   & $-1.075$ & $1.868\cdot 10^{6}$   & $-6.211\cdot 10^{5}$   & $6.518\cdot 10^{4}$& $-2.261\cdot 10^{3}$ \\[-0.25em] \cline{3-9} & & & & & & & &  \\
\hline
\end{tabularx}
\caption{\label{tab7}  Table of fit coefficients analog to Tab.\ \ref{tab0}, but for the Fourier coefficients $\beta_{2,1,1}$ and $\beta_{2,2,1}$.}

\end{table*}

\end{appendix}

\end{document}